# Policy Creation Model for Policy-Based Management in Telecommunications Networks

Carlos A. Astudillo, *Graduate Student Member*, *IEEE,* Adriana M. Gustin, *Graduate Student Member*, *IEEE* and Oscar J. Calderón, *Member*, *IEEE*
New Technologies in Telecommunications R&D Group
Faculty of Electronic and Telecommunication Engineering, University of Cauca
Popayán (Cauca), Colombia
{castudillo, amgustin, ocalderon}@ieee.org

*Abstract*— **Policy-based management (PBM) is being used as technological solution on the managing and controlling complex networks and systems. One of the most important issues involved in the life-cycle of PBM is the policies creation because the future decisions made by the management system depend on this, and therefore, the network behavior. In this paper we present a novel model for creating management policies in telecommunications networks. We propose a model which includes a Policy Creation Process, Actors, Policy Abstraction Levels and a Procedure for Creating Policies. An implementation of the proposed model over the Technology Division at University of Cauca is included.**

*Keywords-Policy Based Management, Policy Creation Process, Telecommunications Networks Management.*

## I. INTRODUCTION

The idea of autonomic management has been proposed as the future of several areas such as Security, Quality of Services (QoS), Resources Allocation but at this moment, it is not a reality yet. So we need some benchmarks for creating policies in current networks which use early PBM solutions for managing traditional network elements. Those elements can be routers, switches, firewalls, PCs, etc.

The currents solutions of PBM systems in academic field use some formals techniques for translating high level policies into the low level ones [1],[2]. Those also use policy languages like Ponder [3]. However, they were created in experimental scenarios and they are not being used by service providers to manage their networks. Nowadays, the early PBM solutions are being used yet, so we proposed a model that can help to create policies in both current and future commercial management systems based on PBM.

This work introduces a model that can be used in a manual way but it is inspired in experimental works where the policies are generated in automatic way, using different techniques and formal languages. It would allow our model to be automatic in the future.

The structure of the paper is as follows. Section 2 introduces Policy Based Management and the principal components of its architecture, Section 3 develops the policy creation model, in Section 4 the model is used in a real scenario and finally, the conclusions are presented.

## II. POLICY-BASED MANAGEMENT

The architecture defined by IETF [4] is at this moment the most popular architecture of Policy Based Management (PBM) for researchers, academics and industries. IETF defines four functional elements. The Fig. 1 shows that architecture and the components are described in the following:

- Policy Management Tool (PMT): Entity used by policy administrators for specifying the policies to be applied in the network.

- Policy Repository (PR): the policies generated from PMT are stored in this entity.

- Policy Enforcement Point (PEP): Element where the policies are applied and enforcement.

- Policy Decision Point (PDP): This is the most important element of the architecture because it makes decisions based on policies stored in PR and communicates them to PEPs.

Protocols as LDAP, SNMP and COPS are used to communicate the entities in the architecture [5].

The policies are introducing into PMT through policy rules that follow the paradigm If "Conditions" then "Actions" From IETF. Policy would have others components such as subject, target and trigger events, also a management system could support only some type of policy; it depends on the PBM implementation [6].

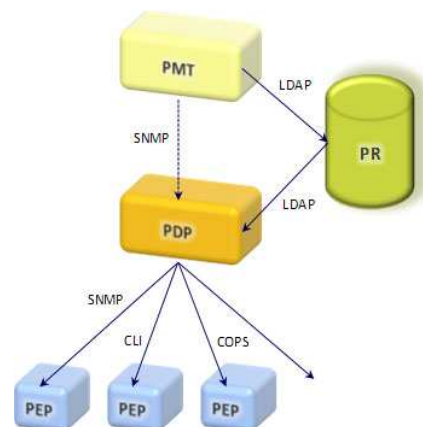

Figure 1. IETF PBM Architecture

## III. POLICY CREATION MODEL

We propose a Policy Creation Model in the context of telecommunications networks which includes Policy Abstraction levels, Policy Creation Process, Actors, and a Procedure for Creating Policies. It is developed in this way:

### A. Policy Abstraction Levels

The policies in a management system can be represented and specified in different ways, providing scalability and flexibility to management systems. The following are the levels proposed by the authors in [7].

*1) High-Level Abstract Policy:* These policies are defined by policy makers. This type of policies includes:

- The business objectives of the company or organization.
- The SLAs defined between providers, providers and their customers, or internally in an organization.
- The needs of those involved in the network, among which are: users/customers, applications, services, providers and network operators.

Such policies are defined in natural language and represent the aims of network behavior and the wishes of the parties involved on it.

Policies at this level are not displayed in the management system and must be refined to Mid-Level Policies to enable them to enter it.

*2) Mid-Level Policy:* These policies are applyed by the administrator in the PMT through policy rules by using Policy Specification Language, Console or Graphical User Interface (GUI).

Policy rule specifies a set of conditions whether they are true, the result is one or more actions. Each rule has a logical statement of the way If "Conditions" Then "Actions", with or without a trigger event, it depends on the PBM implementation.

*3) Low-Level Policy:* Those policies are defined for a particular device, with a specific configuration and represent the lowest level of a policy because they are applied directly to the network element involved in policies (i.e. PEP).

Those policies are built in an understandable and unique format for each device and are converted to this format through the PDP.

Usually at this level policies should not be specified because it loses the objective of policy-based management witch seeks to provide high-level management to a large number of devices at once.

### B. Policy Creation Process

In the following we describe the phases in a policy life-cycle that made up the policy creation process:

*1) High-Level Abstract Policies Definition:* In this phase the management objectives are extracted. those are the requirements for the management system. At this phase is also taked into account the feedback of the cycle which could come from policy monitoring and maintenance. Finally, the High-Level Abstract Policies are writen in a document.

*2) Policy Refinement:* In order to achieve the management objectives, we use a approach based on goals which is inspired in [1] to create operational policies (mid-level policies) in manual way. In our approach, first is defined high-level goals (based on High-Level Abstract Policy) and then are generated sub-goals (operational goals) that can achieve the objectives proposed initially. This refinement makes use of requirements analysis, applications deployed on the network, users of these applications, and resources available.

*3) Definition and Specification of Policy Rules:* In this phase is defined the policies that should be deployed into the management system in order to achieve the management objectives. they are based on low-level goals, conditions and others factors. Those policies can be expressed in a policy specification language (e.g. Ponder, XACML, etc.), or through an information model by using a Graphic User Interface.

### C. Actors

We also identify the actors involved in the policy creation process, these are:

*1) Policy Makers*

This group of people has the task of designing and specifying the High-Level Abstract Policies. This group is integrated by:

*a) Managers:* They know the rules, business objectives and organizational structure.

*b) Network Specialists:* They know the underlay technologies that network can use (e.g. Quality of Service Mechanims, Security and Networking Technologies, etc.).

*c) Administrators:* They Know how to define the management objectives and what the policy system support.

*2) Policy Administrators*

They are responsible for implementing policies on the policy-based management system, to make them maintenance and if required, also carry out maintenance of the necessary mechanisms to enforce these policies. They should also make reports to managers and security, networking and QoS specialists.

*3) Users/Customers*

They are involved in the capture of requirements because they use telecommunications infrastructure, demanding some services or applications.

### D. Procedure for Creating Policies

Identify a generic procedure for the Policy Creation Process in the context of telecommunications networks is essential, because it expedites the process and also ensures the desired results when the policies are implemented to meet the needs for they were created.

The authors have also proposed steps in order to create policies in telecommunications networks management, they are developed in as follows:

*1) Define High-Level Abstract Policies*

The first thing to do in the policy life-cycle is to acquire the needs or management requirements which are captured by Policy Makers. These involve:

- Service Level Agreements (SLA's).
- Business Objectives.
- Users/Customers Requirements.

Taking into account the needs identified and defined previously, the Policy Makers write the high-level abstract policies. They are written in natural language and define the desired behavior of the network in terms understandable by anyone.

*2) Specify High-Level Goals*

It expresses the aims to be achieved by the management system in general way (abstract level). Since we use a goal approach for creating policies, we need to define high-level goals that fulfill the necessities and requirements defined into the High-Level Abstract Policies.

*3) Define Sub-Goals*

Each high-level goal can be refined into low-level ones (sub-goals or operational goals), the decomposition goals can be either conjunctive (∧, AND) or disjunctive (∨, OR). From this, it forms a refinement hierarchy in which the dependencies among the goals of different levels of refinement are based on the manner with which it was decomposed.

The low-level goals should be defined to perform a specific task. Therefore, the refinement process should be done until this condition is gotten.

*4) Generate Strategies*

Strategy is called to the sequence of sub-goals that should be implemented to obtain the high-level goal in the management system.

The strategy can be encoded in one or more mid-level policies depending exclusively on the particular application domain.

*5) Identify Conditions*

The conditions are circumstances that occur in networks. If these conditions are validated true, trigger a set of actions that affect the network and allow reaching the desired state of it. Therefore, the administrator should identify the conditions that must be considered for the specification of a policy.

In this step should be identified all conditions that help each Sub Goal to be achieved in the system taking into account the functionality that the system supports, determining the source and destination of traffic, the type of service and time in which each Sub Goal is executed.

*6) Identify the Subject and Target*

It is necessary to identify the subject and target that will be specified in the final policy rules. The Subject refers to the entity responsible for implementing the actions involved in the policy or the objects that are permitted or prohibited in the actions. The Target refers to the elements affected by the policy actions.

*7) Define the Policies which will be Deployed by the System*

In order to deploy a policy in a management system, it is necessary to define this in a mid-level policy format (see part A). For this, it uses all the elements obtained in the previous steps (e.g. sub goals, conditions, target, and subject).

The policy administrator needs to define the way in which policies will be specified, this depends on the PBM System. The specification could be done through policy specification languages and GUI, enabling policy information to be shared with other entities such as PDPs and PEPs. The following are the main advantages and disadvantages of both options:

The languages provide a consistent definition of policies which facilitates the interpretation of the elements that run them. Some languages are unique to particular implementations, while others may be used independently of the implementation and therefore not tied to a specific framework.

GUIs are the most common ways to express policies in IETF policy framework because they have not a defined language to specify policies. It makes use of graphical user interfaces in which policies are expressed in the form **if** conditions **then** actions, and the managed objects, actions that are executed on these objects and the conditions under actions which they apply, are defined by the information model. This way of expressing policy is simpler and less experience is required by administrators, but its problem is that the identification of conflicts is more difficult.

There are proposals to use the two types of policy specification previously appointed together, taking advantage of each one, like in [8].

## IV. APPLICATION OF THE PROPOSED MODEL

We used the proposed model over a real scenario at the University of Cauca in order to obtain the policies that should be implemented in the management system for achieving all management objectives. In the following we develop the model in this particular scenario and as a result of this is obtained the policies that will be deployed by the system. We used a commercial PBM solution, called Allot Communications NetPolicy, it allows us to manage the network traffic and some routers with policies that are specified in the PMT (called NetXplorer) and enforced in the PEP (called NetEnforcer).

*A. Actors Description*

*1) Policy Makers*

This group of players is made up of:

- Chief of Servers and Services of Internet Area.

- Chief of Infrastructure Area.
- Support Engineer, Infrastructure Area.

*2) Policy Administrators*

This group is made up of Servers and Services Administrators of Internet Area, University of Cauca.

*3) Users*

This group of players is made up of the service users (e.g. Administrative People, Students, Professors, etc.). they at any time may require a particular network behavior, for any application or service.

*B. Define High-Level Abstract Policies*

The aim of the Technology Division is to create policies to manage the internet traffic, because it is necessary to use effectively the Internet links according to the services that it offers. In this sense, the Policy Makers capture the following needs/requirements:

- Web servers, internet access, web applications, mail server and downloads of special hosts must be available at any time and have assigned a high priority.
- VoIP, videoconferencing and streaming services should be prioritized and must ensure adequate bandwidth for their proper functioning.
- Access to sites of institutional interest must always be guaranted.
- Services non-institutional interesting should have low priority and be restricted during working hours.
- Hosts that have special permits should be having limited bandwidth.

Taking into account the needs described above, it is necessary that the system implements policies to manage its main resource: the bandwidth. The Policy Makers identify the following high-level abstract policies:

- Web servers must be accessed at any time and at a good speed, regardless of the amount of traffic which they arise.
- Mails must arrive and depart in a short period of time.
- Access to websites that are related to administrative and academic work of the university will be prioritized and must ensure a bandwidth (e.g. Banks, others Universities and Governmental Sites (i.e. ICETEX, COLCIENCIAS and ICFES).
- Ensure that the FTP server can always be accessed. However, because it is an additional service and not a principal, the traffic should have a medium priority that allows people quickly download files.
- Access to P2P applications should be restricted during working hours.
- The lowest priority should be provided to all accesses with no institutional use (e.g. video and music downloads, etc).
- Guarantee access to the VoIP server with high QoS.
- The videoconference equipments and some videoconference software servers must be guaranteed bandwidth to enable them to function properly.
- Hosts that have special services (i.e. those with public address or NAT) must have a limited bandwidth, with a lower priority than other important services.
- The Internet surfing through the less important web sites for the university should have a low priority.

*C. Specify High-Level Goals*

Based on the previous step that describes the management needs/requirements of Technology Division, now the Policy Makers define the high-level goal G 1-1: "Bandwidth optimized for both incoming and outgoing internet traffic" which represents the main objective that should be achieved by the management system.

*D. Define Sub-Goals*

In this step, the Policy Makers begin to extract the sub-goals (SG), generating a refinement hierarchy. This is done taking into account the management needs/requirements, available resources and information about the management environment.

It is necessary to take into account the services, the resources consumed by each service and the priority to be given to each of these resources, according to the role in the institutional mission at the university.

The figure 2 shows the refinement hierarchy generated for our particular case, where:

**$SG_{2-1}$**: Allow access to the Services and/or Applications.

**$SG_{2-2}$**: Set a Bandwidth for Services and/or applications by defining thresholds (Minimum - Maximum).

**$SG_{2-3}$**: Ensure a Bandwidth for Services and/or Applications.

**$SG_{3-1}$**: Ensure a bandwidth of 256Kbps per incoming connection to mail servers.

**$SG_{3-2}$**: Set the maximum bandwidth allowed per outgoing connections from mail servers.

**$SG_{3-3}$**: Set a minimum bandwidth of 500Kbps for incoming and outgoing connections of computers that have urgent downloads.

**$SG_{3-4}$**: Ensure a bandwidth of 64Kbps for incoming and outgoing connections to VoIP Server.

**$SG_{3-5}$**: Ensure a bandwidth of 384Kbps per connection for incoming and outgoing traffic of videoconference equipments.

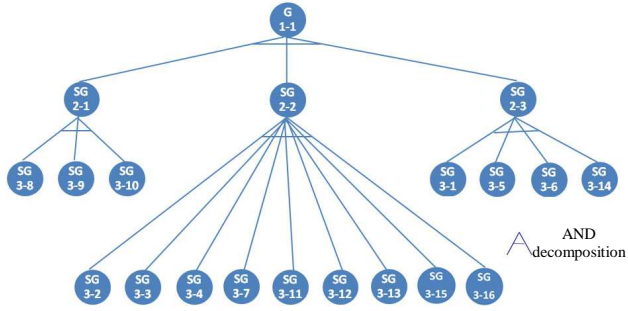

Figure 2. Refinement Hierarchy

**SG$_{3-6}$**: Ensure a bandwidth of 512Kbps for incoming connection to the Web Server.

**SG$_{3-7}$**: Set the maximum bandwidth allowed per connection for outgoing traffic from Web Servers.

**SG$_{3-8}$**: Deny access to RapidShare during working hours.

**SG$_{3-9}$**: Deny access to P2P applications during working hours.

**SG$_{3-10}$**: Allow access to P2P applications during non-working hours.

**SG$_{3-11}$**: Set a minimum bandwidth of 400kbps to internet access for equipment with public IP address.

**SG$_{3-12}$**: Set a maximum bandwidth of 512Kbps for outgoing traffic from FTP server.

**SG$_{3-13}$**: Set a minimum bandwidth of 1024Kbps for outgoing traffic to important web sites.

**SG$_{3-14}$**: Ensure a bandwidth of 300Kbps per connection for outgoing traffic from streaming servers.

**SG$_{3-15}$**: Set a minimum bandwidth of 5Mbps for incoming traffic to proxy servers.

**SG$_{3-16}$**: Set a minimum bandwidth of 300Kbps for outgoing traffic from proxy servers.

Note that Sub Goals SG$_{2-X}$ include a system functionality and the Sub Goals SG$_{3-X}$ define a specific action based on the functions supported by the management system.

### E. Generate Strategies

Since the scenario is not complex enough (has one PEP), the goal graph has only one possibility to achieve the high-level goals (all leads are conjunctive). Therefore, it must meet all low-level goals in order to achieve the high-level ones. The generated strategy is called S1.

S1=SG$_{3-1}$^SG$_{3-2}$^SG$_{3-3}$^SG$_{3-4}$^SG$_{3-5}$^SG$_{3-6}$^SG$_{3-7}$^SG$_{3-8}$^SG$_{3-9}$^SG$_{3-10}$^SG$_{3-11}$^SG$_{3-12}$^SG$_{3-13}$^SG$_{3-14}$^SG$_{3-15}$^SG$_{3-16}$

### F. Identify Conditions

Table I summarizes the conditions identified for each sub goal.

### G. Identify Subject and Target

For this particular case, the subject and target refers to NetEnforcer AC404 because it is the only equipment allowed in the managed domain to enforce policies.

TABLE I. CONDITIONS FOR EACH SUB GOAL

| Sub Goal | Conditions | | | |
|---|---|---|---|---|
| | *Source* | *Destination* | *Service* | *Time* |
| SG$_{3-1}$ | Mail Servers | Any | Mail | Any |
| SG$_{3-2}$ | Any | Mail Server | Mail | Any |
| SG$_{3-3}$ | Hosts with Important Dowloads | Any | All IP | Any |
| SG$_{3-4}$ | VoIP Server | Any | VoIP | Any |
| SG$_{3-5}$ | Videoconference Equiments | Any | All IP | Any |
| SG$_{3-6}$ | Web Servers | Any | All IP | Any |
| SG$_{3-7}$ | Any | Web Server | All IP | Any |
| SG$_{3-8}$ | Any | RapidShare Servers | Web Applications | Working Hours |
| SG$_{3-9}$ | Any | Any | P2P Applications | Working Hours |
| SG$_{3-10}$ | Any | Any | P2P Applications | Non Working Hours |
| SG$_{3-11}$ | Host with NAT or Public IP | Any | All IP | Any |
| SG$_{3-12}$ | FTP Server | Any | FTP | Any |
| SG$_{3-13}$ | Any | Important Web Sites | Web Applications | Any |
| SG$_{3-14}$ | Streaming Servers | Any | All IP | Any |
| SG$_{3-15}$ | Any | Proxy Servers | All IP | Any |
| SG$_{3-16}$ | Proxy Servers | Any | All IP | Any |

### H. Define the Policies which will be Specified in the System

As previously mentioned, these policies are defined in the way If "conditions" Then "actions". They are the rules that must finally be implemented at a policy-based management system to meet the established management requirements.

The following are the defined policies that make up the strategy S1:

**P1** (Based on SG$_{3-1}$): If "The Service is email, the Source Address are mail servers, Destination Address is whichever, anytime" Then "Ensure a bandwidth of 256Kbps to mail servers, and assign a medium priority (6)".

**P2** (Based on SG$_{3-2}$): If "The Service is email, the Source Address is whichever, the Destination Address belongs to one of the mail servers, anytime" Then "Assign a medium priority (6) and give the maximum available capacity to service".

**P3** (Based on SG$_{3-3}$): If "The Source Address belongs to any host of urgent downloads group, Destination Address is whichever, anytime" Then "Assign a minimum bandwidth of 500 Kbps for both inbound and outbound connection to that computer, and give a high priority (9)".

**P4** (Based on SG$_{3-4}$): If "The Source Address and Destination Address is the VoIP server, anytime" Then "Ensure a bandwidth of 64 Kbps for both inbound and outbound connection to that server, and assign a high priority (9)".

**P5** (Based on SG$_{3-5}$): If "The Source Address or Destination Address belongs to any video conference equipments, anytime" Then "Ensure a bandwidth of 384 Kbps for both inbound and outbound connections to these hosts and assign a medium priority (6)".

**P6** (Based on $SG_{3-6}$): If "The Source Address is whichever, the Destination Address belongs to one of the web servers, anytime" Then "Ensure a bandwidth of 512 Kbps for incoming connection to these servers and assign a middle priority (7)".

**P7** (Based on $SG_{3-7}$): If "The Source Address belongs to one of the web servers, Destination Address is whichever, anytime" Then "Give the maximum available bandwidth for outgoing connection from these servers and assign a low priority (4)".

**P8** (Based on $SG_{3-8}$): If "The Source Address is whichever, the Destination Address is one of the RapidShare servers, the Service is a web application, anytime" Then "Discard traffic from and to those servers".

**P9** (Based on $SG_{3-9}$): If "The source address is whichever, the Application is P2P, at working hours" Then "Deny Access to those applications".

**P10** (Based on $SG_{3-10}$): If "The Source Address and Destination Address is whichever, the Application is P2P, at non-working hours" Then "Accept traffic to and from those applications".

**P11** (Based on $SG_{3-11}$): If "The Source Address belongs to one host with NAT or Public IP, Destination Address is whichever, anytime" Then "Assign a maximum bandwidth of 400 Kbps to these hosts and assign a middle priority (7)".

**P12** (Based on $SG_{3-12}$): If "The Source Address belongs to one of the FTP servers, Destination Address is whichever, anytime" Then "Set a maximum bandwidth of 512 Kbps for outgoing traffic from FTP servers and assign a low priority (4)".

**P13** (Based on $SG_{3-13}$): If "The Source Address is whichever, Destination Address is one of the important web sites, anytime" Then "Set a minimum bandwidth of 1024 Kbps".

**P14** (Based on $SG_{3-14}$): If "The Source Address is one of the streaming servers, Destination Address is whichever, anytime" Then "Ensure a bandwidth of 300 Kbps per connection and assign a middle priority (6)".

**P15** (Based on $SG_{3-15}$): If "The Source Address is whichever, the Destination Address is one of the proxy servers, anytime" Then "Set a minimum bandwidth of 5 Mbps and assign a middle priority(7)".

**P16** (Based on $SG_{3-16}$): If "The Source Address is one of the proxy servers, Destination Address is whichever, anytime" Then "Set a minimum bandwidth of 500 Kbps and assign a middle priority (5)".

Those policies are introduced in the management system, mapping the entities involved on them. For example, the group of hosts with urgent downloads is represented in the system by its IPs and the working hours are from 8:00 A.M to 6:00 P.M. Since policies above were created with the system functionality, it can be introduced in the system. The ranges established to determine the priority assigned to each policy are like follow: from 1 to 4, low priority, from 5 to 7, middle priority and finally from 8 to 9, high priority. The ranges are provided by PBM implementation.

## V. CONCLUSION

Policy-Based Management is considered of great interest within the academics, business and researchers because it allows greater flexibility in management operations, regarding the translation of business requirements into specific policy rules.

We identify individuals or actors that are involved in the policy creation process and we clarified the roles of them. We could make an appropriate construction of policies if we had qualified human resources (administrative and technical support) and an effective interaction between them (roles well defined for all actors involved in the policy creation process).

The proposed abstraction levels to manage telecommunications networks exposed that policies can not be seen as a single entity or statically, as in these new environments of dynamic networks and specific requirements for each network user, the policies must be created and implemented at different levels within the management plane of the network.

One of the most important results of this work is the procedure for creating policies, which allows developing through steps each phases involved in this process. Additionally, this procedure is linked with policy abstraction levels and actors, and an application example is developed.

ACKNOWLEDGMENT

The authors would like to thanks Technology Division, University of Cauca for providing the facilities for implementing the Policy Creation Model as well as its staff for their help and advices in special to Jaime Gaviria, Jaime Martinez, Andres Zúñiga, Fabio Fuertes and Eivar Armero. We would also like to thanks Carlos Silva from Cambridge Language Centres and anonymous reviewers for their usefull comments.